\documentclass[prb,twocolumn,showpacs,preprintnumbers,amsmath,amssymb]{revtex4}

\usepackage{graphicx}
\usepackage{dcolumn}
\usepackage{bm}

\begin{document}


\title{Decoupling transition of two coherent vortex arrays within the surface superconductivity state}

\author{Alexey V. Pan$^{1,2}$}
\email{pan@uow.edu.au}
\author{Pablo Esquinazi$^1$}

\affiliation{$^1$Abteilung Supraleitung und Magnetismus,
Institut f\"ur Experimentelle Physik II, \\
Universit\"at Leipzig, Linn\`estra{\ss}e 5, D-04103 Leipzig, Germany \\
$^2$Institute for Superconducting and Electronic Materials,
University of Wollongong, \\ Northfields Avenue, Wollongong, NSW 2522, Australia}

\date{May 9, 2003, Revised: July 21, 2004}

\begin{abstract}
In magnetic fields applied within the angular range of the surface
superconductivity state a magnetically anisotropic layered medium is
created in structurally isotropic, sufficiently thick niobium films.
Surface (Kulik) vortices residing in the superconducting sheaths on both
main film surfaces in tilted fields are shown to undergo a decoupling
transition from a coherent to an independent behavior, similar to the
behavior observed for Giaever transformer. At the
transition a feature in pinning properties is measured, which implies
different pinning for the lattice of surface vortices coherently coupled
through the normal layer and for two decoupled vortex arrays in the
superconducting surface sheaths.
\end{abstract}


\maketitle

\section{Introduction}

The existence of vortices and vortex states within the surface
superconductivity (SC) state has been discussed since the
discovery \cite{ssc} of this phenomenon within the applied
magnetic field range $B_{c2} \le B_a \le B_{c3}$ ($B_{c2}$ and
$B_{c3}$ are the second and surface superconductivity critical
fields). The existence of Abrikosov vortices \cite{Kulik52},
Abrikosov-like state \cite{ziese,pan}, giant (multi-quantum)
vortex state \cite{rod,RS,fink4andexplosive}, and Kulik (surface)
vortices \cite{kulik,interf1,interf2} above $B_{c2}$ were
suggested depending on sample size and shape and applied field
orientation. In this work, we mainly deal with the magnetic
behavior governed by {\em surface vortices} in the superconducting
surface sheath. The structure of a surface vortex, to a large
extent, reproduces the structure of an Abrikosov vortex with the
length equal to the thickness of the superconducting layer
(surface sheath). Fink and Kessinger showed \cite{fink6} that the
thickness of this sheath $d_{sc} \simeq 1.6 \xi(T)$ at $B_a \simeq
B_{c2}$ and approaches $\xi$ at $B_a = B_{c3}$  for a
superconductor with $\kappa = 10$ ($\kappa = \lambda/\xi$ is the
Ginzburg-Landau parameter, $\xi$ and $\lambda$ are the coherence
length and magnetic field penetration depth). In the case of a
relatively thin superconductor, as the films investigated in this
work, the superconducting surface sheaths on the main film
surfaces are separated by a normal layer
\begin{equation}
d_n \simeq d_p - 3.2\xi(T) \,
\label{dn}
\end{equation}
with $d_p$ being the film thickness.

\begin{figure}[t]
\includegraphics[scale=0.45]{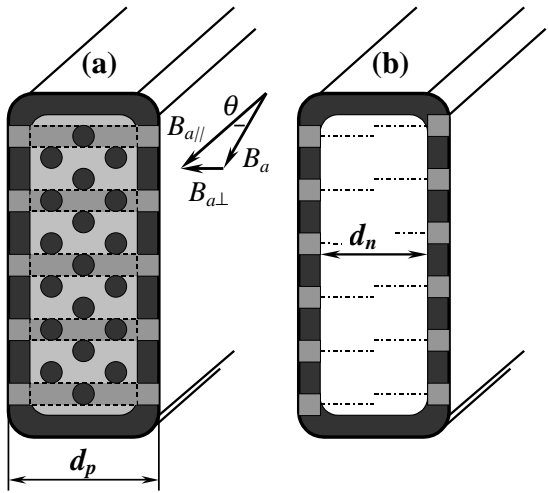}
\vspace{-5.5cm} \caption{\label{model}The vortex arrangements are
schematically shown for the film cross-section in a field $B_a$ applied at
an angle $\theta$ to the  main film surfaces (a) below $B_{c2}$, and (b)
above $B_{c2}$ after the decoupling transition. The black layer near the
circumference of the film denotes the surface SC sheath with the light
gray surface vortices in it. The black circles in (a) imply in-plane
Abrikosov vortex rows \cite{Kulik52,pan} parallel to the surfaces. The dark
grey stripes perpendicular to the surfaces show
the out-of-plane Abrikosov vortices.
Although it is not clear on the picture but it is assumed that vortex lines
of both Abrikosov lattices in (a) do not cross one another.}
\end{figure}

For a certain magnetic field range applied at an angle ($\theta$) to the
surfaces, two independent flux-line lattices (FLLs) can be formed in thin
films and layered systems.
\cite{pan,kulik,interf1,interf2,pan2,phd,blamire,bul,glaz,pannew} This
co-existence is possible due to strong structural or magnetic anisotropy
and to the two components of the applied field. \cite{kulik,bul} In this
case, the out-of-plane field component ($B_{a\perp}$) would be responsible
for the out-of-plane (perpendicular) FLL and the in-plane (parallel)
component ($B_{a||}$) for the in-plane vortex lattice. The co-existence of
two FLLs has experimentally been shown for structurally isotropic films
similar to those investigated in this work \cite{pan,pan2,phd,blamire}.
Therefore, we will hereafter assume that in the films investigated two
co-existing FLLs, perpendicular and parallel to the main film surface, are
present at fields tilted with respect to the surface. Fig.~\ref{model}(a)
schematically shows the two-FLLs structure.

We further assume that at $B_a > B_{c2}$ and over the angular range of
$|\theta| < 40^\circ$, which is the range over which the surface SC state
can be measured:\cite{ziese,pan} (i) the parallel FLL in thick films $d_p
>> \xi$ transforms into a giant vortex above $B_{c2}$.
\cite{RS,fink4andexplosive,phd,pannew} This is a reasonable assumption
because the surface SC state, just like the in-plane Abrikosov vortex
rows, forms due to the $B_{a||}$ component. The shielding super-currents
of the giant vortex flow within the surface sheath. (ii) The perpendicular
FLL forms two arrays of quasi-2D surface vortices residing in the
superconducting sheaths on both main surfaces of the films \cite{kulik}
(Fig.~\ref{model}(b)). The co-existence of the giant vortex and the
surface vortices was discussed in Refs.~\onlinecite{interf1,interf2}. It
was also shown that Kulik vortices can form triangular or square lattice,
depending on the applied field and its orientation $\theta$.\cite{kulik}

Neither theoretical nor experimental work has been published, which shows
any kind of interaction between the two surface vortex arrays on the opposite
surfaces of a thin flat sample in the surface SC state. We expect that the
arrays can behave either coherently or independently,
depending on $d_n$, $d_{sc}$, $B_a$ and $\theta$.
These parameters can affect the coupling force
between the vortices as it was shown for the case of two magnetically
coupled superconducting films (superconducting Giaever transformer \cite{g}) in
fields $B_a < B_{c2}$. \cite{clem75a,clem75b} At the transition between the
coherent and independent regimes a {\em small pinning change} would be
expected. Therefore, one needs a technique sensitive to such small changes
in fields applied {\em nearly parallel} to the film surface. Such fields
are necessary to enable the surface SC state. In this work, we describe
results of mechano-magnetic experiments on niobium (Nb) films of different
thicknesses and provide experimental evidence for the decoupling
transition of the surface vortices within the surface SC state.

\section{Experimental details}

The experimental technique employed in this work is based on
mechanical oscillations of a superconductor attached to a
Vibrating Reed (VR) in an external magnetic field (for a review
see Ref.~\onlinecite{Esquin91} and references therein). The
experimental VR setup employed in this work can be found in
Ref.~\onlinecite{pan}. This technique is very sensitive to
magnetic properties of the superconductor, in particular to the
pinning of vortices, in fields parallel to its largest surface
($\theta = 0^\circ$). The physical reason for this sensitivity is
that the very small field component $\sim B_a \varphi$
perpendicular to the applied field, arising when the
superconductor is tilted by a very small angle (typically $\varphi
< 10^{-5}$~degrees), is shielded by superconducting currents
(generally defined by the pinning of vortices). The shielding
currents cause the external field to curve around the tilted
superconductor. This field distortion leads to an additional line
tension (stiffness) in the system, which is proportional to the
increase of the length of the field lines near the edge of the
tilted superconductor. As a result, the resonance frequency
($\omega$) of the VR with the attached superconducting sample
increases with field. If the shielding supercurrents become
smaller, for example in the vicinity of the upper critical field
or the critical temperature (in general due to a pinning
reduction), the resonance frequency decreases. The damping
($\Gamma$) of the oscillator, which is measured simultaneously
with the resonance frequency, is proportional to the corresponding
energy dissipation occurring in the oscillator (reed plus
superconducting sample) due to vortex movement and the internal
friction of the reed material. The peak in the damping, usually
measured as a function of field or temperature, corresponds to the
vortex depinning line.\cite{ziesereview} However, in the case of
the surface superconductivity (the so-called giant vortex state)
the peak can have a different origin related to the
shielding/pinning properties of the giant vortex (see, for
example, Ref.~\onlinecite{RS} and references therein).

In the case of a {\em thin} conventional superconductor, such as Nb-film,
the behavior of the VR as a function of temperature and field is, to a
large extent, governed by the magnetic properties of the surface of the
superconductor. Therefore, the unique properties of the VR technique
should allow us to detect changes in the behavior of the
surface vortex arrays, whose pinning can influence the shielding property.
Accordingly, the resonance
frequency change $\omega^2(B_a)-\omega^2(0)$ and the damping $\Gamma(B_a)$
measured in the experiment are respectively expected to provide
information on surface vortex pinning and energy dissipation produced
by vortex movement.

The increase in the resonance frequency vanishes as soon as vortex pinning
and the shielding become negligible. Thus, we can measure not only
$B_{c2}$ at $\theta \rightarrow 90^\circ$ and $B_{c3}$ at $\theta
\rightarrow 0^\circ$, but also the angular dependence of the upper
critical field $(B_{uc}(\theta))$.\cite{ziese,pan,phd} Naturally, we
define $B_{uc}(\theta = 0^\circ) \equiv B_{c3}$ and $B_{uc}(\theta =
90^\circ) \equiv B_{c2}$.

Another important feature of our experimental setup is the high
angular resolution of the rotation system.\cite{pan} As a
consequence, the angle $\theta$ of the field with respect to the
main film surface was defined with an accuracy of $\pm 0.01^\circ$
in the vicinity of $0^\circ$ and of $\pm 0.5^\circ$ at $\theta
\gtrsim 3^\circ$. At $\theta \gg 0^\circ$, the angular resolution
is smaller  for these experiments due to a small
$\Delta^2B_{cu}/(\Delta\theta)^2$ at $\theta \neq 0^\circ$ (see
Fig.~5 in Ref.~\onlinecite{pan}). The accuracy near $0^\circ$ is
limited by the sensitivity of the Si-oscillator onto which the
superconducting sample is attached. This Si-oscillator has a
quality factor $Q \sim 10^6$ at the temperatures of the
measurements.

Nb-films of different thicknesses $d_p \simeq 120$~nm (Nb120), 400~nm
(Nb400), and 1200~nm (Nb1200) were investigated in this work. The
polycrystalline films were sputtered onto an oxidized silicon wafer at
room temperature.\cite{films} Superconducting properties of these films
were characterized in earlier works. \cite{pan,pan2,phd} The coherence
length at zero temperature $\xi(0)$ for all the measured samples was
estimated to be $\sim 12.5$~nm ($\xi(T = 5\, {\rm K}) \simeq 19$~nm). We
estimate a Ginzburg-Landau parameter $\lambda/\xi \sim 10$ for the three
films.\cite{pan}

\section{Experimental results}

\begin{figure}[t]
\vspace{2.5cm}
\includegraphics[scale=0.34]{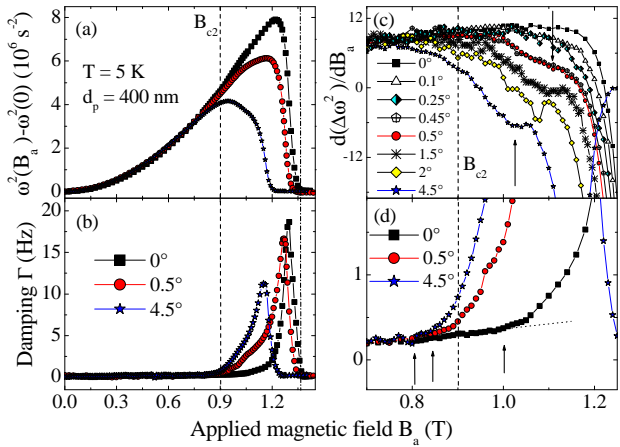}
\vspace{-3.5cm}
\caption{\label{rf400}(a) The resonance frequency change $\Delta \omega^2
\equiv \omega^2(B_a)-\omega^2(0)$, (b) the corresponding damping, (c) the
first derivative of $\Delta \omega^2$ on $B_a$ with  the arrows denoting
the minima ($B_m$), and (d) the enlargement of the damping onset at
$B_{\rm onset}$ which is marked by the arrows for the Nb400-film.
The dashed and dashed-dotted lines mark $B_{c2}$ and $B_{c3}$, respectively.}
\end{figure}

\begin{figure}[t]
\vspace{2.5cm}
\includegraphics[scale=0.34]{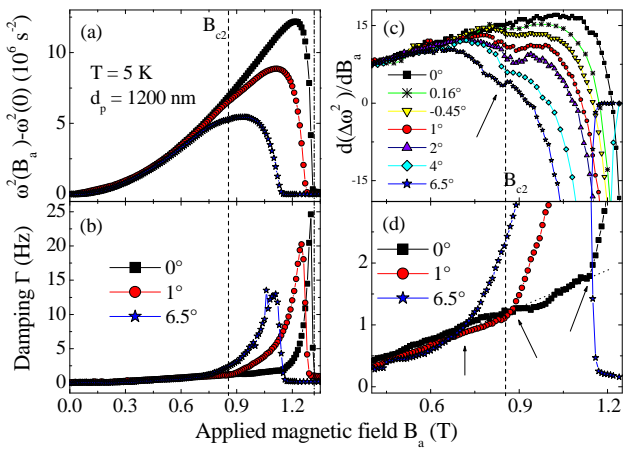}
\vspace{-3.5cm}
\caption{\label{rf1200}The same as in Fig.~\ref{rf400} for the Nb1200-film.}
\end{figure}

Figures~\ref{rf400} and \ref{rf1200} show the measured resonance
frequency change $\omega^2(B_a)-\omega^2(0)$ (a) and the VR
damping $\Gamma(B_a)$ (b) as a function of applied field at
different angles $\theta$ and at fixed temperatures for the Nb400 and
Nb1200-films, respectively. The key-feature in these figures is
the appearance of an unusual non-monotonic behavior at angles
$\theta \ge \theta_m^{on}$, which is best seen as a minimum at a
field we define as $B_m$ in the first derivative of the resonance
frequency (Figs.~\ref{rf400}(c) and \ref{rf1200}(c)). This
``critical" angle $\theta_m^{on}$ is $\simeq 0.45^\circ$ and
$0.16^\circ$ for the Nb400- and Nb1200-films, respectively. $B_m
(\theta)$ for both films is shown in Fig.~\ref{onset}(a). The
angular range of the non-monotonic behavior ($\theta_m^{on} \le
\theta_m < 20^\circ $) is within the range of the surface SC
existence. \cite{pan,phd} In principle, one would tend to observe
the position of the minima ($B_m$) at $B_{c2}$. In this case, the
minima would naturally indicate a change in the shielding property
when the bulk superconductivity collapses and only surface
superconductivity persists. However, the minima do not coincide
with the experimentally measured $B_{c2}$ (marked by the dashed
lines in Figs.~\ref{rf400}, \ref{rf1200}, and \ref{onset}).
Instead, the $B_m$-behavior is more complex being angular
dependent. We argue below that this behavior can be explained as
{\it decoupling of the coherent arrays of the surface vortices},
which undergo a decoupling transition from coherent behavior at
$B_a < B_m$ to independent behavior at $B_a > B_m$. This
transition is promoted by the magnetically anisotropic medium
created in the films in fields within $B_{c2} \le B_a < B_{c3}$
applied nearly parallel to the surface. Taking into account the
thickness of the films, the coupling between the coherent surface
vortex pairs is of {\em magnetic} nature.

\section{Discussion}

\subsection{Angle dependence of $B_m$}

\begin{figure}[t]
\vspace{-0.8cm}
\begin{center}\leavevmode
\includegraphics[scale=0.4]{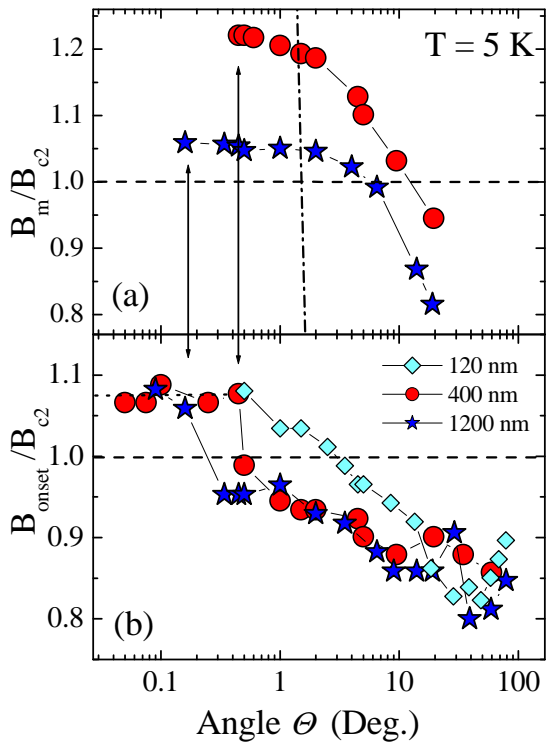}
\caption{\label{onset}(a) Normalized field $B_m$ at which minima are
observed in the resonance frequency change, and (b) damping onset as a
function of angle. The arrows mark $\theta_m^{on}$ in (a) coinciding with
the step-like feature in (b) for the Nb400- and Nb1200-films. The
dash-dotted, nearly vertical line in (a) shows $B_{cr}(\theta)$ obtained from
Eq.~(\protect\ref{a}) assuming a fixed value for the intervortex distance
$a_0 = a^{cr}_0 = 2 \lambda(5$~K$)= 380$~nm. The dotted line in (b) shows
the plateau at small angles and an additional experimental point measured
at $\theta = 0^\circ$ for the films, which cannot be shown in the
logarithmic scale.}
\end{center}
\end{figure}

At $\theta < \theta_m^{on}$, $B_{a\perp}$ is too small to induce a
large density of surface vortices in the surface sheaths. This is likely to
imply that the magnetic behavior in this angular range is overwhelmingly
governed by the giant vortex surface shielding current. Therefore,
the VR signal is not sensitive enough to reveal a possible decoupling
transition for only a few surface vortices up to $B_{cu}$.
As $\theta$ becomes larger
(Fig.~\ref{onset}(a)), $B_{a\perp}$ increases and more {\em coupled} surface
vortices are created. These vortices
start interacting within each sheath at a characteristic crossover
field given by \cite{clem75b}
\begin{equation}
B_{cr} = \frac{2\Phi_0}{\sqrt{3}(a^{cr}_0)^2} \, ,
\label{23d} \end{equation}
where $\Phi_0$ is the flux quantum and $a^{cr}_0$ is the
intervortex distance at the crossover. As soon as intervortex
interaction between surface vortices within one sheath becomes
stronger than the coupling force, the decoupling between the
coherent pairs of surface vortices takes place, forming two
independent arrays of (2D-like) vortex lattices, one in each
surface sheath. The magnetic coupling of the surface vortices is
weak due to the relatively large distance $d_n$. Thus, relatively
weak in-plane vortex-vortex interaction should be enough to
decouple the coherent behavior. A sufficiently strong intervortex
interaction for a decoupling would occur at an intervortex
distance $a^{cr}_0 \sim 2\lambda(T)$. If we assume that the
intervortex spacing for the triangular surface vortex lattice is
\cite{kulik}
\begin{equation}
a_0 \simeq (2\Phi_0/\sqrt{3}B_a\sin\theta)^{0.5} \,,
\label{a} \end{equation}
one finds that at $\theta_m^{on}$, $a_0 \simeq 0.5\,\mu$m for the
Nb400-film and $a_0 \simeq 1\,\mu$m for the Nb1200-film. Thus, the
intervortex spacing is  of the order of $2\lambda(T = 5 {\rm K})
\sim 0.38\,\mu$m at $\theta_m^{on}$.

Let us assume that the decoupling transition occurs when the
perpendicular component of the applied field $B_{a\perp} \equiv
B_a \sin\theta$ is equal to $B_{cr}$. Then, the dash-dotted line
in Fig.~\ref{onset}(a) shows $B_{cr}(\theta) \propto (a_0^2
\sin\theta)^{-1}$ expected from Eq.~(\ref{a}) with the fixed $a_0
= a^{cr}_0 = 2\lambda(T = 5 {\rm K}) = 380$~nm for the Nb1200- and
Nb400-films. As can be seen, Eq.~(\ref{a}) with the fixed $a_0$
does not describe the $B_m(\theta)$ behavior over the entire
decoupling line except at the angle $\sim 1.5^\circ$ at which the
calculated curve crosses the corresponding experimental
$B_m$-lines. The disagreement between the experimental curves and
Eq.~(\ref{a}) should actually be expected. Indeed, in the case of
Eq.~(\ref{a}) only one parameter -- $a_0(B_{a\perp})$, -- affecting the
decoupling, changes with field (or angle). Whereas in our case
there are at least four variables affecting the coupling:
$a_0(B_{a\perp})$, $d_{sc}(B_a||)$, $d_n(d_{sc})$ and surface
vortex pinning. The behavior of the $B_m(\theta)$ curves in
Fig.~\ref{onset}(a) can  be explained by four major factors, which
influence the above mentioned variables: (i) the nature of the
decoupling which occurs within the surface SC state and with an
enormously large interlayer spacing $d_n$ and small $d_{sc}$, (ii)
surface roughness of the measured films, (iii) surface vortex
pinning which is not accounted for in Eq.~(\ref{a}), and (iv)
angular dependence of the upper critical field $B_{cu}$.

The {\em first} factor responsible for the $B_m$ behavior below
the crossing point is particularly well described on the example
of the thicker film (Nb1200) with {\em much weaker coupling} (due
to the larger $d_n$) than that in the Nb400-film. For the
Nb1200-film the $\theta$-independent plateau is observed for $B_m$
at $\theta \le 2^\circ$. In this range the decoupling is driven
by the reduction of $d_{sc}$ above $B_{c2}$ with increasing
$B_a$.\cite{fink6} The smaller $d_{sc}$ (the larger $d_n$) leads
to a reduction of the coupling and pinning for both surface
vortices of all coupled pairs. As soon as the decoupling threshold
is reached, the surface vortices from each pair are likely to be
dragged apart by shielding currents incoherently oscillating on
opposite surface sheaths. As $\theta$ approaches the crossing
point at $\sim B_{cr}$, $B_m$ starts to curve downwards being also
affected by intervortex interactions. A similar, but stronger
effect experiences the thinner film (Fig.~\ref{onset}(a)). The
stronger angular dependence below the crossing point is likely
observed because the decoupling threshold is higher than for the
thicker film. Hence, to reach the decoupling higher fields have to
be applied (Fig.~\ref{onset}(a)), which result in larger surface
vortex populations and, consequently, in stronger intervortex
interactions.

The {\em second} factor can have some influence in the vicinity of the
crossing point. The values of $a_0$, calculated for ``ideal" film
surfaces, are likely to be underestimated due to the surface roughness
present in real films.\cite{hs} The surface roughness model implies that
even if $\theta = 0^\circ$, the flux would intercept some localized areas
of the rough surface. Thus, the rougher the surface, the more surface
vortices are expected to populate the sheaths in applied fields nearly
parallel to the surface. The Nb1200-film was found to have a
larger value of the root mean square surface roughness (6.6~nm) than the
Nb400-film (5.3~nm).\cite{phd} This result can contribute to the fact
that the decoupling has been observed starting from a smaller $\theta_m^{on}$
for the Nb1200-film than for the thinner film.
Apparently, at larger angles the
surface roughness factor becomes less significant.

The {\em third} factor is responsible
for the disagreement at larger angles. The pinning experienced by
the surface vortices
\cite{interf2,hs,surfpin}, which we neglected to a large extent in the above
consideration, is likely to modify significantly Eq.~(\ref{a})
derived by assuming pinning-free environment.\cite{kulik}

The {\em fourth} factor can also influence  $B_m$ in particular at larger
angles, since, for example, $B_{cu}$ at $5^\circ$ is about 10\% smaller
than $B_{c3}$.\cite{pan,phd} This can affect the thickness of the surface
sheath $d_{sc}$, pinning and shielding properties, and, therefore,
the decoupling.

In Fig.~\ref{onset}(a) one sees that increasing $\theta$,
$B_m$ approaches $B_{c2}$.
It may seem surprising that the feature attributed
to the decoupling in the
surface SC state still exists {\it below} $B_{c2}$. However, it was shown
in a number of theoretical \cite{fink4andexplosive,gv,f} and experimental
works \cite{phd,pannew,tsind} that a giant vortex state within a
superconducting surface sheath can be nucleated at sufficiently high
fields below $B_{c2}$. In this case, the magnetic anisotropy (the layered
structure) of the surface SC state can also be preserved below $B_{c2}$.
In addition, the superconducting order parameter within the $d_n$-layer is
substantially reduced due to a large number of densely packed in-plane
Abrikosov vortices.\cite{f} In this case the magnetic anisotropy
(layered structure) is effectively maintained below $B_{c2}$.
We stress that the decoupling is
observed only within the angular range of the surface SC state existence
($0 \le \theta \le 40^\circ$)\cite{pan,phd} defined as $B_{cu} > B_{c2}$.
Therefore, the decoupling appears to be a realistic scenario below
$B_{c2}$, as well.

Summarizing, the decoupling behavior in Fig.~\ref{onset} can be described
as follows. At $\theta < \theta_m^{on}$, the giant vortex shielding
overwhelmingly dominates, so that the possible decoupling of only very few
surface vortices cannot be detected by the VR technique:
neither the measured resonance
frequency change nor damping show an unusual
behavior up to the vicinity of $B_{c3}$ as if there were no transition at
$B_{c2}$. At $\theta \ge \theta_m^{on}$ the decoupling occurs at
$B_m(\theta)$. At fields below the decoupling pinning and shielding
properties behave in a usual way as described for the VR technique.\cite{pan,Esquin91,ziesereview} As the field further increases
the coupling force between the sufficiently large amount of surface vortex
pairs becomes too small to prevent the decoupling. The decoupling is likely driven by two different mechanisms below and above the crossing point:\\
(i) Below the crossing point, as the coupling forces become too small due to
the $d_{sc}$ reduction with increasing field,
the independent surface vortices become
more mobile due to incoherent oscillation of the shielding currents on the
opposite film's surfaces. As the result, the shielding properties slightly
weakens (resonance frequency), and the dissipation notably onsets (damping).
However, the mobility of the vortices is expected to be incomplete due to
the arising pinning of individual surface vortices. \\
(ii) Above the crossing point, this scenario is further complicated by
an additional parameter: intervortex interaction, which assists in the
decoupling process. In this case, the mobility of the vortices would be
restricted by a collective process which arises from interplay between
vortex-vortex interaction and pinning of the surface vortices.\\
In both cases, the shielding would be slightly weakened at the
decoupling and partially regained after pinning independent
2D-like surface vortex arrays in the superconducting surface
sheaths. Experimentally, this behavior has produced the observed
minima in the first derivative of the resonant frequency change
(Figs.~\ref{rf400}(c) and \ref{rf1200}(c)) and the apparent
enhancement of the damping at $B_{\rm onset}$ (see the arrows in
Figs.~\ref{rf400}(d) and \ref{rf1200}(d)). Note that in
Fig.~\ref{onset}(b), the $B_{\rm onset}(\theta)$ dependences
for the Nb400 and Nb1200-films show
step-like features at $\theta_m^{on}$. These steps coincide
precisely with the appearance of $B_m$ in the resonance frequency
change for both films.

\subsection{Coupling force}

\begin{figure}[t]
\vspace{2.5cm}
\begin{center}\leavevmode
\includegraphics[scale=0.34]{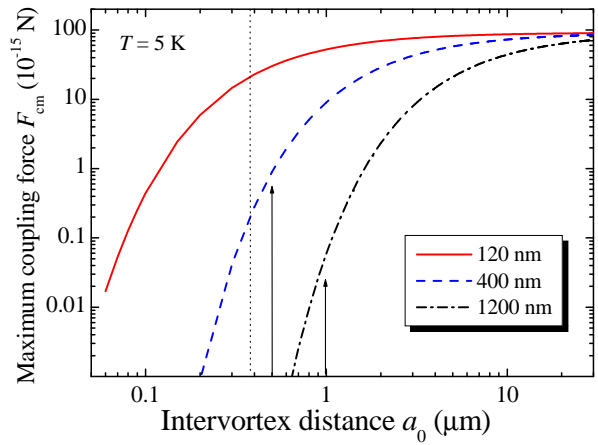}
\vspace{-3.0cm} \caption{\label{force}The maximum coupling force
($F_{cm}$) between two surface vortex arrays on the opposite film
surfaces as a function of the intervortex distance ($a_0(B_{a\perp}$))
between the surface vortices (Eq.~\ref{fcm}). The dotted line indicates
$a_0 = 2 \lambda = 0.38$~$\mu$m at $T = 5$~K.}
\end{center}
\end{figure}

For simplicity, assuming that the thickness of the superconducting
surface layer  $d_{sc} \simeq 1.6\xi$ (in fact,
$d_{sc}(B_{a\perp})$ \cite{fink6}), the maximum coupling force
($F_{cm}$) at high-flux density regime ($B > B_{cr}$) is given by
\cite{clem75b}
\begin{equation}
F_{cm} = \frac{3 \Phi_0^2 a_0^2}{32 \pi^4 \mu_0  \lambda^4} \left[1 -
\exp(-\frac{2\pi}{a_0}d_{sc}) \right ]^2 \exp(-\frac{2\pi}{a_0}d_n)\,,
\label{fcm} \end{equation}
provided that  $d_{sc}/\lambda \ll 1$, $d_n \ll \lambda^2/d_{sc}$,
and $\xi \ll a_0/(2\pi) \ll \lambda$, where $\mu_0$ is the
permeability of free space. However, we should note that some of
the actual conditions for the Nb1200-film at $T = 5$~K are
slightly softer than the given above: $d_n < \lambda^2/d_{sc}$ and
$a_0/(2\pi) < \lambda$. We believe that it should be acceptable
for our estimate, especially taking into account that the main
condition of thin superconducting layers $d_{sc}/\lambda \ll 1$ is
fulfilled.\cite{clem75b} In addition, we also note that $\lambda$
becomes larger and $d_{sc}$  smaller with increasing field,
reinforcing the applicability of the corresponding inequalities.

In Fig.~\ref{force}, $F_{cm}$ as a function of $a_0(B_{a\perp})$ is shown
for all the films. At the decoupling crossover ($\theta_m^{on}$) indicated
by the arrows, $F_{cm} \simeq 8.9 \times 10^{-16}$~N and $5.9 \times
10^{-17}$~N for the Nb400- and Nb1200-films, respectively.

Taking into account the trend for thinner films to produce the
decoupling onset ($\theta_m^{on}$) at larger angles and stronger
magnetic fields (Fig.~\ref{onset}(a)), the expected decoupling
onset for the Nb120-film would be at or slightly below $a_0 \simeq
2\lambda$ (the dotted line in Fig.~\ref{force}). In this region,
$F_{cm}$ is {\em much larger} for the Nb120-film than that for the
thicker films. Importantly, it is nearly within the region where
$F_{cm}$ is nearly independent on $B$. Therefore, to observe the
decoupling transition a much larger $B_{a\perp}$ (smaller $a_0$),
implying larger $\theta$, should be applied in order to reduce
$F_{cm}$ and to increase the intervortex interaction. However, a
minimum in the resonance frequency corresponding to the decoupling
was not observed for this film, nor  the step-like feature in the
behavior of $B_{\rm onset}(\theta)$ (Fig.~\ref{onset}(b)).
Instead, $B_{\rm onset}(\theta)/B_{c2}$ is clearly larger than for
the thicker films. This behavior indicates stronger shielding and
pinning, which remain unaffected by the decoupling but affected by
the critical field dependence $B_{cu}(\theta)$ only.

\section{Conclusion}

In summary, from VR experiments on thin Nb films we have obtained
evidence indicating that for sufficiently thick films the surface
vortices in the surface SC state undergo a decoupling transition.
At small fields/angles the aligned vortices are coupled through
the normal layer $d_n$ exhibiting a coherent 3D-like vortex
lattice behavior. At larger fields/angles the surface vortices
decouple forming two independent vortex arrays (2D-like behavior)
in the superconducting surface sheaths. In films with $d_p
\lesssim \lambda$, the coupling between the aligned surface
vortices appears to be too strong so that the experimental
observation  is not possible with the VR technique. By comparison,
we note that the loss of the 3D coherence in a lattice of aligned
pancake vortices in layered high-temperature superconductors was
explained in terms of a melting phase transition from the 3D
vortex pinning state to the regime of independently pinned 2D
vortex lattices for Josephson \cite{glaz} and magnetic
\cite{glazkes} couplings between the superconducting layers.

\begin{acknowledgments}
We would like to thank C. Assmann (PTB, Berlin) for providing us the films
and R. H\"ohne for the support during the measurements.
\end{acknowledgments}


\begin{thebibliography}{99}
\bibitem{ssc} D. Saint-James and P. G. de Gennes,
Phys. Lett., {\bf 7}, 306 (1963); C. F. Hempsted and Y. B. Kim,
Phys. Rev. Lett. {\bf 12}, 145 (1964); W. J. Tomasch and
A. S. Joseph, Phys. Rev. Lett. {\bf 12}, 148 (1964).
\bibitem{Kulik52} I. O. Kulik, Zh. Eksp. Teor. Fiz. {\bf 52} (Rus.), 1632 (1967) [JETP {\bf 25}, 1085 (1967)].
\bibitem{ziese}M. Ziese, P. Esquinazi, S. Knappe, and H. Koch, J. Low Temp. Phys. {\bf 103}, 71 (1996).
\bibitem{pan}A. V. Pan, M. Ziese, R. H\"ohne, P. Esquinazi, S. Knappe, and H. Koch, Physica C {\bf 301}, 72 (1998).
\bibitem{rod} P. R. Doidge and K. Sik-Hung, Phys. Lett. {\bf 12}, 82 (1964).
\bibitem{RS}R. W. Rollins and J. Silox, Phys. Rev. {\bf 155}, 404 (1967).
\bibitem{fink4andexplosive}H. J. Fink, Phys. Rev. Lett. {\bf 16}, 447 (1966);
M. Ghinovker, I. Shapiro, and B. Ya. Shapiro, Phys. Rev. B {\bf 59}, 9514 (1999);
V. Bruyndoncx, J. G. Rodrigo, T. Puig, L. Van Look, V. V. Moshchalkov, and
R. Jonckheere, Phys. Rev. B {\bf 60}, 4285 (1999).
\bibitem{kulik}I. O. Kulik, Zh. Eksp. Teor. Fiz. {\bf 55}, 889 (1968) [JETP {\bf 28}, 461 (1969)].
\bibitem{interf1}P. Monceau, D. Saint-James, and G. Waysand, Phys. Rev B {\bf 12}, 3673 (1975).
\bibitem{interf2}P. Mathieu, B. Pla\c cais, and Y. Simon, Phys. Rev. B {\bf 48}, 7376 (1993).
\bibitem{fink6}H. J. Fink, R. D. Kessinger, Phys. Rev. {\bf 140}, A1937 (1965).
\bibitem{bul}L. N. Bulaevskii, M. Ledvij, and V. G. Kogan, Phys. Rev. B {\bf 46}, 366 (1992).
\bibitem{glaz}L. I. Glazman and A. E. Koshelev, Phys. Rev. B {\bf 43}, 2835 (1991).
\bibitem{pan2}A. V. Pan, R. H\"ohne, M. Ziese, P. Esquinazi, and C. Assmann,
in {\em Physics and Materials Science of Vortex
States, Flux Pinning and Dynamics}, NATO Science Series {\bf 356}, Kluwer A.P. (Dordrecht, 1999), p. 545.
\bibitem{phd}A. V. Pan, {\em PhD thesis}, University of Leipzig (2000), unpublished.
\bibitem{blamire}M. G. Blamire, C. H. Marrows, N. A. Stelmashenko, and J. E. Evetts, Phys. Rev. B {\bf 67}, 014508 (2003).
\bibitem{pannew}A. V. Pan, R. H\"ohne, and P. Esquinazi, Physica B {\bf 329-333}, 1377 (2003).
\bibitem{g}I. Giaever, Phys. Rev. Lett. {\bf 15}, 825 (1965);
\bibitem{clem75a}J. R. Clem, Phys. Rev. B {\bf 12}, 1742 (1975);
\bibitem{clem75b}J. W. Ekin and J. R. Clem, Phys. Rev. B {\bf 12}, 1753 (1975).
\bibitem{Esquin91}P. Esquinazi, J. Low Temp. Phys. {\bf 85}, 139 (1991).
\bibitem{ziesereview}M. Ziese, P. Esquinazi and
H. F. Braun, Superconductor Science and Technol. {\bf 7}, 869 (1994).
\bibitem{films}S. Knappe, C. Elster, and H. Koch, J. Vac. Sci. Technol. A {\bf 15}, 2158 (1997).
\bibitem{hs}H. R. Hart and P. S. Swartz, Phys. Rev. {\bf 156}, 403 (1967).
\bibitem{surfpin}H. J. Fink, L. J. Barnes, Phys. Rev. Lett. {\bf 15}, 792 (1965); P. S. Swartz, H. R. Hart, Phys. Rev. {\bf 156}, 412 (1967).
\bibitem{gv}L. J. Barnes and H. J. Fink, Phys. Rev. {\bf 149}, 186 (1966); H.J. Fink and A.G. Presson, Phys. Rev. {\bf 151}, 219 (1966).
\bibitem{f}H. J. Fink, Phys. Rev. Lett. {\bf 21}, 853 (1965).
\bibitem{tsind}M. I. Tsindlekht and I. Felner, Physica B {\bf 329-333}, 1371 (2003).
\bibitem{glazkes}A. E. Koshelev, P. H. Kes, Phys. Rev. B {\bf 48}, 6539 (1993).

\end{thebibliography}

\end{document}